# Chip-integrated plasmonic cavity-enhanced single nitrogen-vacancy center emission

Hamidreza Siampour,* Shailesh Kumar and Sergey I. Bozhevolnyi

Centre for Nano Optics, University of Southern Denmark, Campusvej 55, DK-5230 Odense M, Denmark.

*Corresponding author. E-mail: hasa@mci.sdu.dk

High temporal stability and spin dynamics of individual nitrogen–vacancy (NV) centers in diamond crystals make them one of the most promising quantum emitters operating at room temperature. We demonstrate a chip-integrated cavity-coupled emission into propagating surface plasmon polariton (SPP) modes narrowing NV center's broad emission bandwidth with enhanced coupling efficiency. The cavity resonator consists of two distributed Bragg mirrors that are built at opposite sides of the coupled NV emitter and are integrated with a dielectric-loaded SPP waveguide (DLSPPW), using electron-beam lithography of hydrogen silsesquioxane resist deposited on silver-coated silicon substrates. A quality factor of ~ 70 for the cavity (full width at half maximum ~ 10 nm) with full tunability of the resonance wavelength is demonstrated. An up to 42-fold decay rate enhancement of the spontaneous emission at the cavity resonance is achieved, indicating high DLSPPW mode confinement.

Chip-scale, bright and photostable single-photon sources are critical components for quantum cryptography and quantum information processing.[1,2] Colour centers in diamond are very promising candidates among different emitters that have been considered for quantum optical applications.[2-10] The most prominent emitter in diamond is the nitrogen vacancy (NV) center, in which the negatively charged state forms a spin triplet in the orbital ground state, and allows for optical initialization and readout at room temperature.[11] In addition, NV center is a stable single-photon source at room temperature. However, the resonant optical emission of an NV center at a wavelength of 637 nm (zero-phonon line, ZPL) is weak, being less than 4% of total emission even at cryogenic temperatures. The resonant emission is accompanied by a broad phonon sideband ranging from ~ 600 nm up to 800 nm at room temperature. For some quantum optical applications only the photons emitted in the ZPL are useful.[12-15] To enhance and channel the emission into a narrow band, the environment of an emitter can be engineered.[16-22] Regarding the emission enhancement at the ZPL, photonic and plasmonic cavities have been employed.[21-29] Photonic cavities are diffraction limited, and high quality factors (Q) of the cavity required to reach high Purcell effects ultimately limit the rate of emission.[23,25] Plasmonic cavities, on the other hand, feature only moderate Q due to absorption losses, but small volumes can be achieved.[23,25] Plasmonic cavities can be used for channelling the emission into a waveguide as well, as has been demonstrated with an NV center coupled to a plasmonic cavity fabricated around a chemically grown silver nanowire.[30] It is however tedious and time consuming to build a circuitry using chemically grown silver nanowires.[31]

In this work, we demonstrate a compact plasmonic configuration based on a broadband NV quantum emitter resulting in a narrow-band single-photon source with colour-selective emission enhancement. The idea is to combine the surface plasmon polariton (SPP) confinement with relatively low insertion loss by employing a hybrid plasmonic-photonic waveguide-cavity design and to achieve thereby a significant enhancement in the decay rate of NV spontaneous emission at the cavity resonance.

Deterministic placement of emitters is crucial to the integration of colour centers in diamond with quantum optical networks.[32-40] Recently, we demonstrated a top-down nanofabrication technique for precise on-chip positioning of plasmonic waveguide components with respect to single-photon emitters.[41] Using this technique we are able of determining the in-plane NV position within ~30 nm, which provides the required length scale for accurate placement of an NV emitter inside a single-mode cavity resonator operating in the visible spectral range. Here, we employ the top-down fabrication of dielectric loaded SPP waveguides (DLSPPWs), which support well-confined modes with relatively low propagation loss as compared with other metallic plasmonic platforms,[42] and demonstrate compact cavities that exhibit filtering abilities of distributed Bragg reflectors (DBRs) and modify NV-center emission. We define DLSPPW-based DBR-cavities by patterning hydrogen silsesquioxane (HSQ) e-beam resist deposited on silver film (Fig. 1). The DLSPPW-based DBR supports a fundamental SPP mode whose effective refractive index varies through quarter wave stack layers of air/HSQ, creating a number of repeated pairs of low/high refractive index needed to reflect the SPP mode.

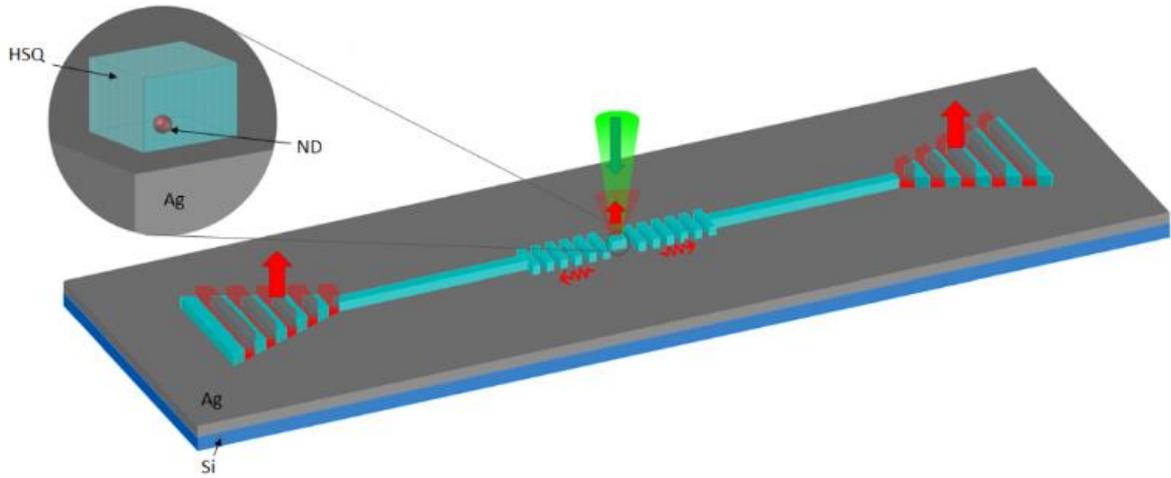

**Fig. 1** Schematic of chip-integrated cavity-coupled NV emitter in nanodiamond (ND).

The HSQ DLSPPW ridges have refractive index of 1.41, and dimensions of 250 nm in width and 180 nm in height. Fig 2a shows a top view image of the Bragg cavity structure. The transverse ridges of the DBR are 140 nm in width, 750 nm in length, and repeated with a period of Λ = 325 nm. The length of the HSQ rectangle equals one-half of the SPP wavelength ($\lambda_{SPP}/2$) in longitudinal direction, and the gap between HSQ rectangle and DBR is 185 nm. Finite difference time domain (FDTD) simulations (Lumerical Solutions, Inc.) indicate a stopband dip in the DBR transmission (Fig. 2c) and a resonant feature when two DBRs are built together to form a cavity (Fig. Fig. 2b and Fig. 2e). We used Palik's handbook of optical constants for modelling of silver refractive index.[43] In the experiment, a silicon wafer coated with a silver film of 250 nm thickness (via thermal evaporation in a vacuum pressure of 2E-7 Torr, and a rate of 4 nm/s). The HSQ e-beam resist (DOW CORNING XR-1541-006) is then spin coated (1200 rpm, 1 min) to make a 180 nm film on silver layer and then the DBR and cavity structures are fabricated using electron-beam lithography (EBL). Experimental characterization of fabricated structures clearly shows the stopband transmission for DBR (Fig. 2d) and exhibits a resonance peak inside the stopband in the cavity transmission, as predicted by the simulations (Fig. 2f).

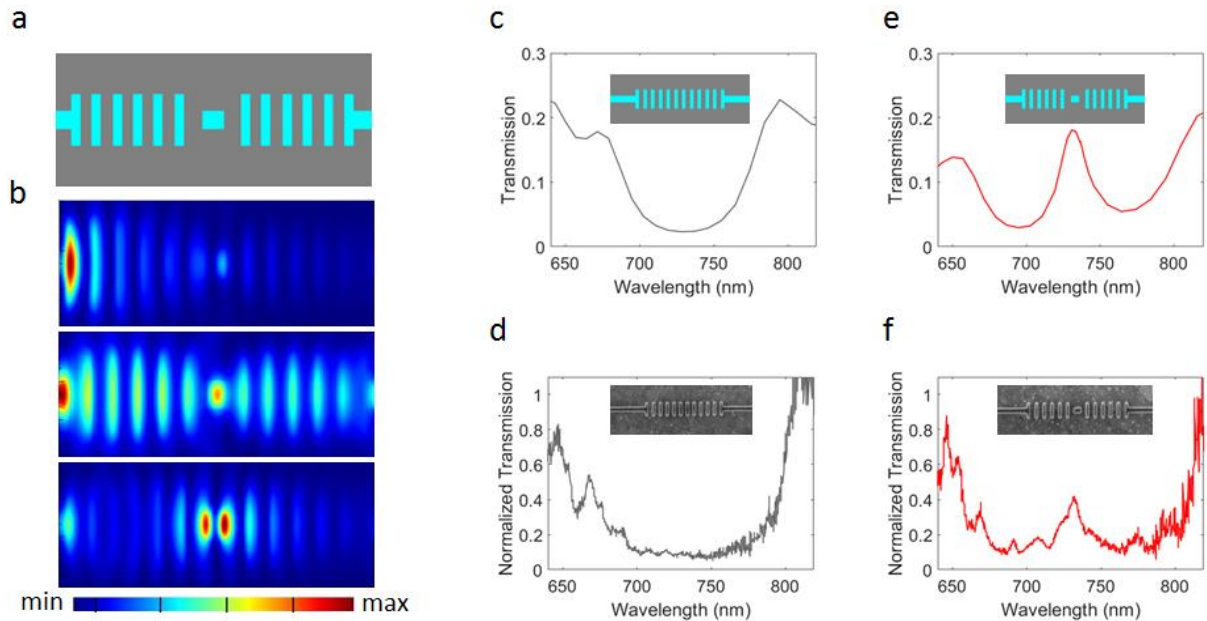

**Fig. 2** (a) Schematic of Bragg grating cavity (top view). (b) Simulated E-field intensity profile of the cavity at wavelengths inside the stopband at 760 nm (top), outside the stop band at 790 nm (middle), and on resonance at 730 nm (bottom). (c) Simulated transmission of the Bragg grating stopband. (d) Experimental result for the fabricated DBR. (e) Simulated transmission of the Bragg cavity. (f) Experimental result for the fabricated cavity.

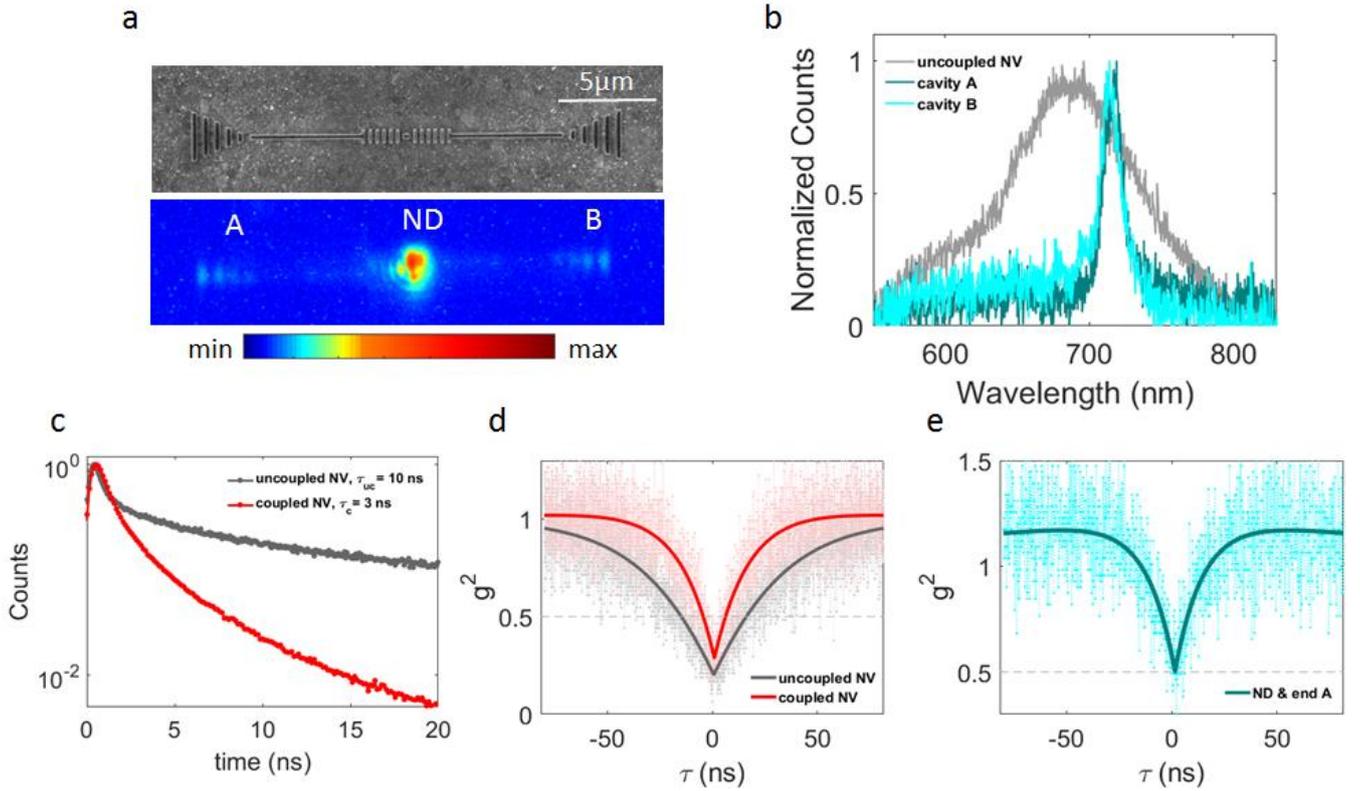

**Fig. 3** (a) SEM image of the fabricated device (top), and CCD camera image of the whole structure when the ND is excited with a continuous wave (532nm) laser (bottom). (b) Spectrum taken from uncoupled NV center (grey), and from the coupled NV at out-of-cavity ends of A (dark green) and B (cyan). (c) Lifetime of the NV-center taken before (grey) and after (red) coupling. (d) Autocorrelation of NV center before (grey) and after (red) coupling. (e) Cross-correlation between out-of-cavity end A and NV center in ND.

NV-centers can be incorporated into the cavity resonators whose resonant wavelength can be selected by appropriately tuning the grating period of DBRs. In the experiment, a silicon wafer coated with a silver film of 250 nm thickness, on which gold markers are made, and subsequently, nanodiamonds (Microdiamant MSY 0-50 nm GAF) are spin coated. The sample with dispersed nanodiamonds is characterized by scanning in a fluorescence confocal microscope and lifetime, spectrum and autocorrelation measurements are taken for NV-centers contained in nanodiamonds. The HSQ e-beam resist is then spin coated and the waveguide-cavity structure is fabricated using EBL onto the nanodiamond, which is found to be a single-photon emitter. An SEM image of the fabricated waveguide-cavity system is illustrated in Fig. 3a (top). Charge-coupled device (CCD) camera image shows the coupling of the emitter to the DLSPPW structure, and subsequent emission from the gratings at the two ends (Fig. 3a, bottom). The antibunching dip in the second-order autocorrelation function of the NV-center is observed both before (grey) and after (red) fabrication of the waveguide (Fig 3d), indicating a single-photon emission ($g^2(0)<0.5$). Cross-correlation between the SPP-coupled emission at the out-of-cavity end A, and the emission collected directly from the NV indicates that the outcoupled fluorescence at the end originates from the same NV center (Fig. 3e). The experimental correlation data are fitted using a well-established model.[44]

Figure 3b shows the emission spectrum of the NV-center taken before (grey line) and after coupling (red line), as well as the spectrum of the out-coupled light from the grating ends of A and B (out-of-cavity ends). The fluorescence spectra from the ends show significant modifications that can be related to the transmission spectrum features of the device. This gives a Q factor of ~70 for the fabricated cavity (~10 nm full width at half maximum, FWHM). The emission intensity at the cavity resonance, $I_r$, is notably higher (by ~ 2.5 times) than the intensity outside of the stopband $I_{out}$. Compared with the transmitted intensity on resonance, $T_r$, measured by the supercontinuum laser, and that outside of the stop band, $T_{out}$, this gives a 6±1 fold radiative decay rate enhancement due to the cavity on resonance ($\Gamma_r$) using $\Gamma_r = (I_r/I_{out})(T_{out}/T_r)$, where $I_r$, and $I_{out}$ are normalized with the baseline fluorescence emission of the NV-center.[30]

In Figure 2c, lifetime of the NV-center before and after coupling is presented. A lifetime reduction (from ~10 ns to ~3 ns) is observed with a single exponential tail fitting of the measured data. On average, the lifetime is decreased by a factor of ~2.5±0.5 that, in addition to the 2-fold reduction due to the silver surface, gives a factor of ~5±1 for the overall Purcell enhancement of a broadband NV emitter coupled to the waveguide ($\Gamma_{tot}/\Gamma_0$). This broadband enhancement combined with the narrowband enhancement results in the overall decay rate enhancement ($\Gamma = \Gamma_r\Gamma_{tot}/\Gamma_0$) being as high as 42 at the cavity resonance peak.

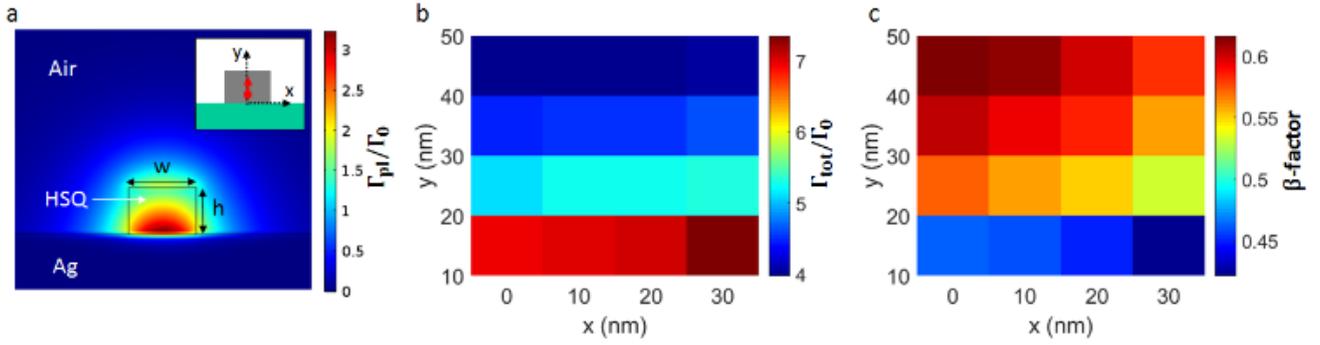

**Fig 4.** (a) Simulated plasmonic decay rate ($\Gamma_{pl}/\Gamma_0$) distribution profile for the DLSPPW coupled NV center. Inset shows the cross section of the NV-center inside the DLSPP waveguide. (b) Distribution profile of the total decay rate ($\Gamma_{tot}/\Gamma_0$) for a random distribution of NV-center inside a ND, and (c) the corresponding β-factor ($\Gamma_{pl}/\Gamma_{tot}$), where each colored square represents the central value of the corresponding in-plane NV position.

In simulations, we decompose the overall decay rate on cavity resonance into two components associated with the longitudinal and transverse confinements. The SPP propagation length ($L_{SPP}$) at the cavity resonance wavelength can be obtained from the relation $L_{SPP} = Q\lambda_{SPP}/2\pi$, in which $\lambda_{SPP}$ is the effective SPP wavelength.[30] We calculate Q factor from FDTD simulations, and $\lambda_{SPP}$ from the effective mode index using finite-element modeling (FEM) method,[45] and thereby $L_{SPP}$ is deduced to be ~ 7.5 μm. This length is less than the value of 20±5 μm estimated for the DLSPPW propagation length,[41] due to the presence of the quarter wave stack in the longitudinal direction of the proposed cavity structure. The longitudinal enhancement ($\Gamma_{long}$) of the plasmonic cavity is then given by $L_{SPP}/L_{eff}$, where $L_{eff}$ is the effective cavity length and can be obtained from the effective mode volume (~ 0.05 $\lambda_{SPP}^3$) and the effective area of the SPP mode (~0.025 $\lambda_{SPP}^2$).[30] This gives the longitudinal enhancement to be 6±1 (the number of roundtrips of a single plasmon in the cavity).

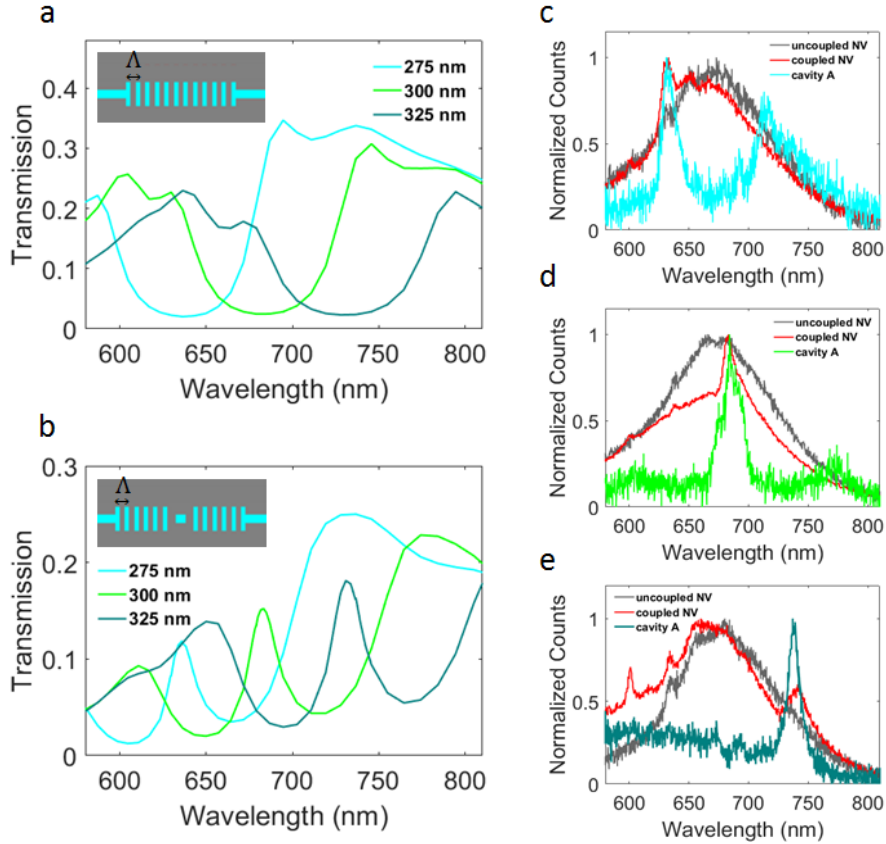

**Fig. 5** Tunable narrowband Bragg grating cavity. (a) Simulated transmission of stopband, and (b) cavity. Experimental results for the cavity-coupled NV⁻ emitter at ZPL (c), at NV⁻ emission peak (d), and into the zero-phonon line of SiV centers (e).

We calculate the enhancement associated with the transverse confinement ($\Gamma_{\text{trans}}$) by simulating the total decay rate ($\Gamma_{\text{tot}}/\Gamma_0$) as explained elsewhere[41] and obtain a value in the range of 4-7.5 for the transverse enhancement using $\Gamma_{\text{trans}} = \Gamma_{\text{tot}}/(\Gamma_0)$, corresponding to the height and lateral position of NV-center (Fig. 4b). Combining these two factors ($\Gamma = \Gamma_{long}\Gamma_{trans}$), we end up with an up to 52-fold overall decay rate enhancement at the cavity resonance. Simulated plasmonic decay rate enhancement ($\Gamma_{\text{pl}}/\Gamma_0$, Fig. 4a) and total decay rate ($\Gamma_{\text{tot}}/\Gamma_0$, Fig. 4b) indicate an up to 63% coupling efficiency (β-factor, Fig. 4c) for the NV-DLSPPW hybrid system, which is in good agreement with the measured value β ~ 58%, as previously reported.[41]

Full tunability of the stopband and cavity over the entire emission range of NV-center is ensured by varying the periodicity of the DBR as predicted in simulated transmissions (Fig. 5a and 5b). We design the quarter wave stack period to have resonance in our desired wavelengths including the zero-phonon line of NV⁻ ($\lambda$=637 nm, Fig. c), NV⁻ emission peak (($\lambda$=680 nm, Fig. d), and the zero-phonon line of silicon-vacancy (SiV) centers ($\lambda$=738 nm, Fig. e). Emission into the ZPL is important for photon-mediated entanglement of internal quantum states of multiple emitters. The proposed cavity-coupled system gives us an opportunity to enhance the emission properties of other colour centers in diamond, and in particular narrow-band germanium-vacancy (GeV, $\lambda_{ZPL}$=602 nm) and SiV centers to achieve even larger Purcell-enhancement.

In conclusion, we have presented a DLSPPW-based cavity coupled to a diamond-based NV emitter, achieving an up to 42-fold decay rate enhancement of spontaneous emission at the cavity resonance. The plasmonic configuration described features enhanced total decay rate and improved spectral purity of the coupled NV emitter, promising thereby potential applications in on-chip realization of quantum-optical networks.

## Supporting Information

Electronic supplementary information (ESI) available: Optical characterization of cavities, influence of NV-center position on the Purcell factor, and distribution of fluorescence lifetimes for single-photon NV emitters. See DOI:10.1039/c7nr05675c

## Acknowledgements

The authors acknowledge financial support from the European Research Council, Grant 341054 (PLAQNAP).

## References

1. J. L. O'Brien, A. Furusawa and J. Vuckovic, *Nat Photon*, 2009, **3**, 687-695.
2. I. Aharonovich, D. Englund and M. Toth, *Nat Photon*, 2016, **10**, 631-641.
3. T. B. Hoang, G. M. Akselrod and M. H. Mikkelsen, *Nano Letters*, 2016, **16**, 270-275.
4. X. Guo, C.-l. Zou, C. Schuck, H. Jung, R. Cheng and H. X. Tang, *Light Sci Appl.*, 2017, **6**, e16249.
5. C. Xiong, B. Bell and J. Eggleton Benjamin, *Nanophotonics*, 2016, **5**, 427-439.
6. P. Reineck, M. Capelli, D. W. M. Lau, J. Jeske, M. R. Field, T. Ohshima, A. D. Greentree and B. C. Gibson, *Nanoscale*, 2017, **9**, 497-502.
7. B. Lienhard, T. Schröder, S. Mouradian, F. Dolde, T. T. Tran, I. Aharonovich and D. Englund, *Optica*, 2016, **3**, 768-774.
8. J. J. Cadusch, E. Panchenko, N. Kirkwood, T. D. James, B. C. Gibson, K. J. Webb, P. Mulvaney and A. Roberts, *Nanoscale*, 2015, **7**, 13816-13821.
9. S. K. H. Andersen, S. Kumar and S. I. Bozhevolnyi, *Nano Letters*, 2017, **17**, 3889-3895.
10. T. Schröder, S. L. Mouradian, J. Zheng, M. E. Trusheim, M. Walsh, E. H. Chen, L. Li, I. Bayn and D. Englund, *J. Opt. Soc. Am. B*, 2016, **33**, B65-B83.
11. M. W. Doherty, N. B. Manson, P. Delaney, F. Jelezko, J. Wrachtrup and L. C. L. Hollenberg, *Physics Reports*, 2013, **528**, 1-45.
12. E. Togan, Y. Chu, A. S. Trifonov, L. Jiang, J. Maze, L. Childress, M. V. G. Dutt, A. S. Sorensen, P. R. Hemmer, A. S. Zibrov and M. D. Lukin, *Nature*, 2010, **466**, 730-734.
13. H. Bernien, B. Hensen, W. Pfaff, G. Koolstra, M. S. Blok, L. Robledo, T. H. Taminiau, M. Markham, D. J. Twitchen, L. Childress and R. Hanson, *Nature*, 2013, **497**, 86-90.
14. W. Pfaff, B. J. Hensen, H. Bernien, S. B. van Dam, M. S. Blok, T. H. Taminiau, M. J. Tiggelman, R. N. Schouten, M. Markham, D. J. Twitchen and R. Hanson, *Science*, 2014, **345**, 532-535.
15. B. Hensen, H. Bernien, A. E. Dreau, A. Reiserer, N. Kalb, M. S. Blok, J. Ruitenberg, R. F. L. Vermeulen, R. N. Schouten, C. Abellan, W. Amaya, V. Pruneri, M. W. Mitchell, M. Markham, D. J. Twitchen, D. Elkouss, S. Wehner, T. H. Taminiau and R. Hanson, *Nature*, 2015, **526**, 682-686.
16. A. V. Akimov, A. Mukherjee, C. L. Yu, D. E. Chang, A. S. Zibrov, P. R. Hemmer, H. Park and M. D. Lukin, *Nature*, 2007, **450**, 402-406.
17. R. Kolesov, B. Grotz, G. Balasubramanian, R. J. Stohr, A. A. L. Nicolet, P. R. Hemmer, F. Jelezko and J. Wrachtrup, *Nat Phys*, 2009, **5**, 470-474.
18. A. Huck, S. Kumar, A. Shakoor and U. L. Andersen, *Physical Review Letters*, 2011, **106**, 096801.
19. S. Kumar, A. Huck and U. L. Andersen, *Nano Letters*, 2013, **13**, 1221-1225.
20. E. Bermudez-Urena, C. Gonzalez-Ballestero, M. Geiselmann, R. Marty, I. P. Radko, T. Holmgaard, Y. Alaverdyan, E. Moreno, F. J. Garcia-Vidal, S. I. Bozhevolnyi and R. Quidant, *Nat Commun*, 2015, **6**, 7883.
21. A. Faraon, P. E. Barclay, C. Santori, K.-M. C. Fu and R. G. Beausoleil, *Nat Photon*, 2011, **5**, 301-305.
22. A. Faraon, C. Santori, Z. Huang, V. M. Acosta and R. G. Beausoleil, *Physical Review Letters*, 2012, **109**, 033604.
23. S. I. Bozhevolnyi and J. B. Khurgin, *Optica*, 2016, **3**, 1418-1421.
24. K. Beha, H. Fedder, M. Wolfer, M. C. Becker, P. Siyushev, M. Jamali, A. Batalov, C. Hinz, J. Hees, L. Kirste, H. Obloh, E. Gheeraert, B. Naydenov, I. Jakobi, F. Dolde, S. Pezzagna, D.

Twittchen, M. Markham, D. Dregely, H. Giessen, J. Meijer, F. Jelezko, C. E. Nebel, R. Bratschitsch, A. Leitenstorfer and J. Wrachtrup, *Beilstein Journal of Nanotechnology*, 2012, **3**, 895-908.
25. S. I. Bozhevolnyi and J. B. Khurgin, *Nat Photon*, 2017, **11**, 398-400.
26. H. Kaupp, C. Deutsch, H.-C. Chang, J. Reichel, T. W. Hänsch and D. Hunger, *Physical Review A*, 2013, **88**, 053812.
27. S. Johnson, P. R. Dolan, T. Grange, A. A. P. Trichet, G. Hornecker, Y. C. Chen, L. Weng, G. M. Hughes, A. A. R. Watt, A. Auffèves and J. M. Smith, *New Journal of Physics*, 2015, **17**, 122003.
28. H. Kaupp, T. Hümmer, M. Mader, B. Schlederer, J. Benedikter, P. Haeusser, H.-C. Chang, H. Fedder, T. W. Hänsch and D. Hunger, *Physical Review Applied*, 2016, **6**, 054010.
29. R. Albrecht, A. Bommer, C. Deutsch, J. Reichel and C. Becher, *Physical Review Letters*, 2013, **110**, 243602.
30. N. P. de Leon, B. J. Shields, C. L. Yu, D. E. Englund, A. V. Akimov, M. D. Lukin and H. Park, *Physical Review Letters*, 2012, **108**, 226803.
31. S. Kumar, N. I. Kristiansen, A. Huck and U. L. Andersen, *Nano Letters*, 2014, **14**, 663-669.
32. W. Pfaff, A. Vos and R. Hanson, *Journal of Applied Physics*, 2013, **113**, 024310.
33. S. Johnson, P. R. Dolan and J. M. Smith, *Progress in Quantum Electronics*, 2017, **55**, 129-165.
34. D. M. Toyli, C. D. Weis, G. D. Fuchs, T. Schenkel and D. D. Awschalom, *Nano Letters*, 2010, **10**, 3168-3172.
35. J. M. H. Birgit, M. B. Thomas, T. C. Jennifer, S. H. Jonathan, H. Sungkun, B. Irfan, Y. Amir, D. L. Mikhail and L. Marko, *New Journal of Physics*, 2011, **13**, 045004.
36. I. E. Zadeh, A. W. Elshaari, K. D. Jöns, A. Fognini, D. Dalacu, P. J. Poole, M. E. Reimer and V. Zwiller, *Nano Letters*, 2016, **16**, 2289-2294.
37. Y.-C. Chen, P. S. Salter, S. Knauer, L. Weng, A. C. Frangeskou, C. J. Stephen, S. N. Ishmael, P. R. Dolan, S. Johnson, B. L. Green, G. W. Morley, M. E. Newton, J. G. Rarity, M. J. Booth and J. M. Smith, *Nat Photon*, 2017, **11**, 77-80.
38. M. K. Bhaskar, D. D. Sukachev, A. Sipahigil, R. E. Evans, M. J. Burek, C. T. Nguyen, L. J. Rogers, P. Siyushev, M. H. Metsch, H. Park, F. Jelezko, M. Lončar and M. D. Lukin, *Physical Review Letters*, 2017, **118**, 223603.
39. A. Sipahigil, R. E. Evans, D. D. Sukachev, M. J. Burek, J. Borregaard, M. K. Bhaskar, C. T. Nguyen, J. L. Pacheco, H. A. Atikian, C. Meuwly, R. M. Camacho, F. Jelezko, E. Bielejec, H. Park, M. Lončar and M. D. Lukin, *Science*, 2016, **354**, 847-850.
40. A. Lohrmann, T. J. Karle, V. K. Sewani, A. Laucht, M. Bosi, M. Negri, S. Castelletto, S. Prawer, J. C. McCallum and B. C. Johnson, *ACS Photonics*, 2017, **4**, 462-468.
41. H. Siampour, S. Kumar and S. I. Bozhevolnyi, *ACS Photonics*, 2017, **4**, 1879-1884.
42. A. Kumar, J. Gosciniak, V. S. Volkov, S. Papaioannou, D. Kalavrouziotis, K. Vyrsokinos, J.-C. Weeber, K. Hassan, L. Markey, A. Dereux, T. Tekin, M. Waldow, D. Apostolopoulos, H. Avramopoulos, N. Pleros and S. I. Bozhevolnyi, *Laser & Photonics Reviews*, 2013, **7**, 938-951.
43. E. Palik, *Handbook of Optical Constants of Solids*, Academic Press, 1st edn., 1985.
44. C. Kurtsiefer, S. Mayer, P. Zarda and H. Weinfurter, *Physical Review Letters*, 2000, **85**, 290-293.
45. Y. Chen, T. R. Nielsen, N. Gregersen, P. Lodahl and J. Mørk, *Phys. Rev. B*, 2010, **81**, 125431.